\documentclass[structabstract]{aa}

\usepackage{graphicx}
\usepackage{rotating}
\usepackage{txfonts}
\usepackage{natbib} 
\usepackage{lineno}
\usepackage{ulem}

\bibpunct{(}{)}{;}{a}{}{,} 

\begin{document}
\title{Catching the radio flare in CTA\,102}
\subtitle{I. Light curve analysis}

\author{C. M. Fromm
\inst{1}, M. Perucho\inst{2,3},  E. Ros\inst{2,1}, T. Savolainen\inst{1}, A. P. Lobanov\inst{1}, J. A. Zensus\inst{1}, 
M. F. Aller\inst{4}, H. D. Aller\inst{4}, M.\,A. Gurwell\inst{5} \and A. L\"ahteenm\"aki\inst{6}}
\institute{Max-Planck-Institut f\"ur Radioastronomie, Auf dem H\"ugel 69, D-53121 Bonn, Germany\
\email{cfromm@mpifr.de}
\and Departament d'Astronomia i Astrof\'\i sica, Universitat de Val\`encia, Dr. Moliner 50, E-46100 Burjassot, Val\`encia, Spain\
\and Departament de Matem\`atiques per a l'Economia i l'Empresa, Universitat de Val\`encia, Av. Tarongers s/n, E-46022 Val\`encia, Spain\
\and Department of Astronomy, University of Michigan, Dennison Building, Ann Arbor, MI, 48109 USA\
\and Harvard-Smithsonian Center for Astrophysics, 60 Garden Street, Cambridge, MA, 02138 USA\
\and Aalto University, Mets\"ahovi Radio Observatory, FI-02540 Kylm\"al\"a, Finland }


\abstract
   {The blazar CTA\,102 (z=1.037) underwent a historical radio outburst in April 2006. This event offered a unique chance to study 
the physical properties of the jet.}
   {We used multifrequency radio and $\mathrm{mm}$ observations to analyze the evolution of the spectral parameters during the flare as a test of the shock-in-jet model under these extreme conditions.}
   {For the analysis of the flare we took into account that the flaring spectrum is superimposed on a quiescent spectrum. We 
reconstructed the latter from archival data and fitted a synchrotron self-absorbed distribution of emission. The uncertainties of the 
derived spectral parameters were calculated using Monte Carlo simulations. The spectral evolution is modeled by the shock-in-jet model, 
and the derived results are discussed in the context of a geometrical model (varying viewing angle) and shock-shock interaction}
   {The evolution of the flare in the turnover frequency-turnover flux density ($\nu_m$-$S_m$) plane shows 
a double peak structure. The nature of this evolution is dicussed in the frame of shock-in-jet models. 
We discard the generation of the double peak structure in the 
$\nu_m$-$S_m$ plane purely based on geometrical changes (variation of the Doppler factor). The detailed modeling of the spectral 
evolution favors a shock-shock interaction as a possible physical mechanism behind the deviations from the standard shock-in-jet 
model.}
   {}

\titlerunning{CTA\,102 light curve analysis}
\authorrunning{C. M. Fromm et al.}

\maketitle
\section{Introduction}
The blazar CTA\,102 (B2230+114) has a redshift of $z=1.037$ \citep{Hewitt:1989p1143}. It is classified as a highly polarized 
quasar (HPQ) with a linear optical polarization above 3\% \citep{VeronCetty:2003p1264} and a {V-band} magnitude of 17.33. The source was 
observed for the first time by \citet{Harris:1960p1135}. \citet{Kardashev:1964p1156} reported on possible signals from an 
extraterrestrial civilization coming from CTA\,102. \citet{Sholomitskii:1965p1247} found the first variation in flux density for a 
radio source. Later observations identified CTA\,102 as a quasar.

Since that time CTA\,102 has been the target of numerous observations at different wavelengths. Besides the aforementioned 
variation in the radio flux density, CTA\,102 also shows variation in the optical band. \citet{Pica:1988p1239} reported a variation 
range of 1.14\,mag around an average value of 17.66\,mag {in 14 years, and an increase of 1.04\,mag within two days in 1978, which is so far the most 
significant outburst.} {CTA\,102 has been monitored since 1986 within the $\mathrm{cm}$-observations of the Mets\"ahovi telescope. The strongest radio flare since the beginning of the monitoring took place around 1997, and a nearly simultaneous outburst in the optical R-band was observed with the Nordic Optical Telescope on La Palma \citep{Tornikoski:1999p1255}.}
The source has been detected in the $\gamma$-ray regime by the telescopes \textit{CGRO}/EGRET and \textit{Fermi}/LAT 
with a luminosity, $L_{\gamma}=5\times10^{47}$erg/s, defining CTA\,102 as a $\gamma$-bright source 
\citep{Nolan:1993p1211,Abdo:2009p2201}. {Merlin and VLA observations at 2\,$\mathrm{GHz}$, 5\,$\mathrm{GHz}$ and 15\,$\mathrm{GHz}$ revealed the kpc-scale structure of CTA\,102 which consists of a central core and two faint lobes.} 
\citep{Spencer:1989p1250}. The brighter lobe has a flux density of $170\,\mathrm{mJy}$ at a distance of $1.6\,\mathrm{arcsec}$ from 
the core at position angle (P. A.) of $143^\circ$ (measured from north through east). The other lobe, with a flux density of 
$75\,\mathrm{mJy}$, is located $1\,\mathrm{arcsec}$ from the center at P. A. $-43^\circ$. The spectral indices {between $2\,\mathrm{GHz}$ and $5\,\mathrm{GHz}$} of the lobes are 
$-0.7$ for the bright and $-0.3$ for the other one. 

\begin{figure*} 
\centering 
\includegraphics[width=17cm]{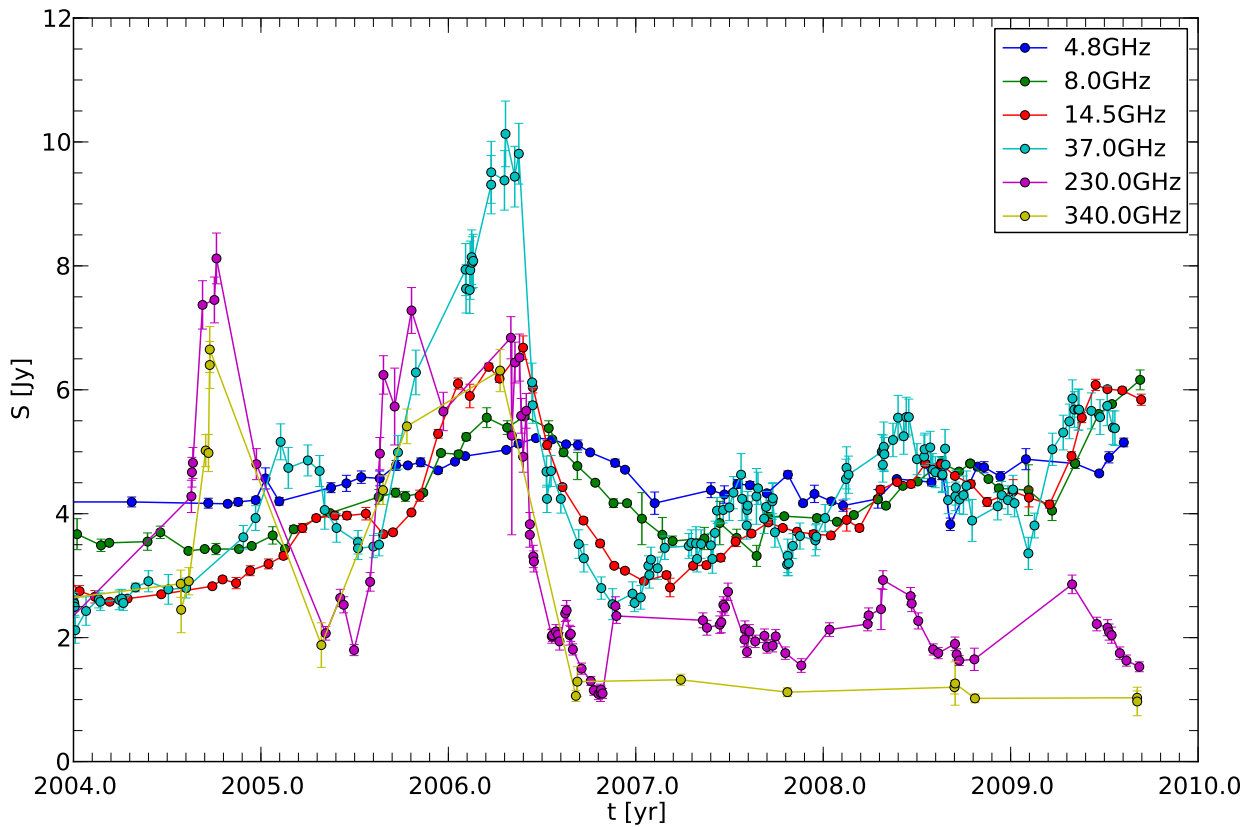} 
\caption{Radio$-$mm light curves for CTA\,102, centered around the 2006 radio flare.} 
\label{lightcurve} 
\end{figure*}

High-resolution VLBI observations at 1.4\,GHz and 5\,GHz resolved the central object into three components and a diffuse 
tail bending to the southeast. These observations provide and upper limit around  $10\,c$ (0.5\,mas/yr) for the superluminal 
motion of the components \citep{Baath:1988p1132, Wehrle:1989p40}. Several observations at different frequencies (for example
at 326\,$\mathrm{MHz}$) confirmed the elongation of the source to the southeast \citep{Altschuler:1995p1128}.

Within the framework of the VLBA 2\,cm-Survey \citep[e.g.,][]{Zensus:2002p2202} and the MOJAVE\footnote{http://www.physics.purdue.edu/MOJAVE} program (Monitoring of Jets in Active galactic 
nuclei with VLBA Experiments) \citep[e.g.,][]{Lister:2009p90} CTA\,102 has been monitored regularly, beginning in mid 1995. These intensive 
observations deliver a detailed picture on the morphology and kinematics of this source at 15\,$\mathrm{GHz}$. Kinematic analysis show apparent velocities of the features in the jet between 0.7\,c and 15.40\,c 
\citep{Lister:2009p8}. A multifrequency VLBI study including data at $90$\,GHz, $43$\,GHz and $22$\,GHz was reported by 
\citet{Rantakyro:2003p1240}. The results from the multifrequency VLBI observations were combined with the continuum monitoring performed 
at single-dish observatories at $22$\,GHz, $37$\,GHz, $90$\,GHz and $230$\,GHz. Within this multifrequency data set (November 1992 
until June 1998), a major flare in CTA\,102 around 1997 was confirmed. The authors could conclude that this event was connected 
to the ejection of a new jet feature. The same was noted by \citet{Savolainen:2002p2410}.
\citet{Jorstad:2005p64} and \citet{Hovatta:2009p7} found Lorentz factors, $\Gamma$, of $17$ and $15$, respectively, and Doppler 
factors, $\delta$, between $15$ and $22$ associated to this ejection. The 2006 radio flare in CTA\,102 has been observed at $\mathrm{cm}-\mathrm{mm}$ total flux density and multifrequency VLBI observations \citep{Fromm:2009diploma}. The results of the the multifrequency VLBI observations will be presented in second paper.  

In this work, we concentrate on the analysis of the $\mathrm{cm}-\mathrm{mm}$ light curves. 
The organization of the paper is the following. In Sect.~\ref{obs} we present the radio/mm light curves during the 2006 
flare and perform the spectral analysis. {The theoretical background 
and the fitting technique are introduced in Sect.~\ref{theory}. The results of this analysis are shown in Sect.~\ref{results}. and are applied to the 2006 flare in CTA\,102 in Sect.~\ref{appmod}}. In Sect.~\ref{disc} we discuss the different models that can explain the observations.\\

{Throughout this paper we define the optically thick spectral index as $\alpha_{t}>0$ and the convention used for the optically thin spectral index is $S\propto \nu^{+\alpha}$, where $\alpha_{0}<0$. The latter can be derived from the spectral slope, $s$, via the relation $\alpha_{0}=-(s-1)/2$.}

\begin{figure*} 
\centering 
\includegraphics[width=17cm]{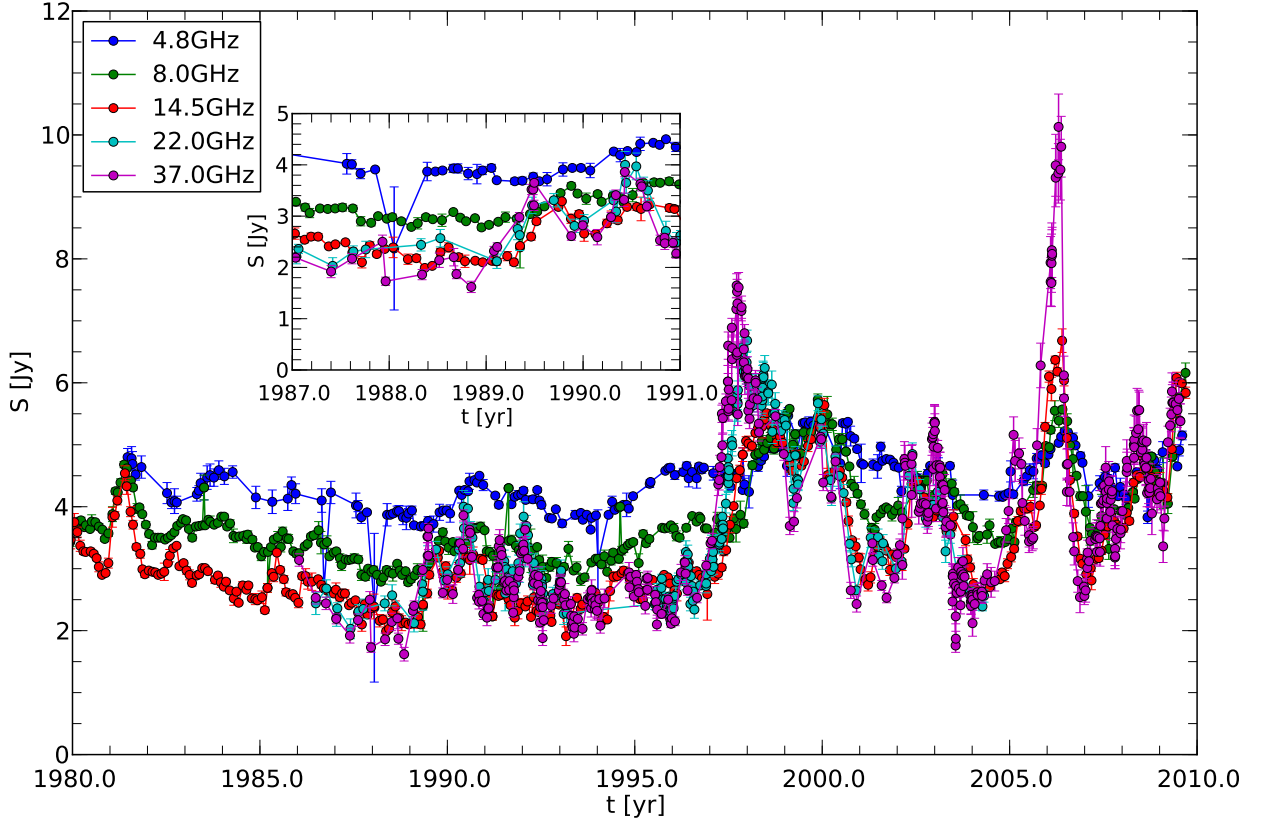} 
\caption{Archival low frequency light curves for CTA\,102. See insert plot for the absolute quiescent state.} 
\label{archival} 
\end{figure*}

\section{Observations: cm$-$mm light curves}
\label{obs}
For our analysis we focused on the radio flare around April 2006 and used observations spanning from 4.8\,GHz to 340\,GHz  
(see Fig.~\ref{lightcurve}). The observations have been carried out by the Radio Observatory of the University of Michigan (UMRAO), the Mets\"ahovi Radio Observatory, and the Submillimeter Array (SMA). The average {sampling time intervals} and flux density uncertainties are presented in Table \ref{obsstat}.

\begin{table} 
\caption{Average time sampling and average flux density uncertainties for the used light curves}  
\label{obsstat}
\centering  
\begin{tabular}{c c c c} 
\hline\hline
$\nu$ [$\mathrm{GHz}$]& Observatory & $\left<t_{\mathrm{obs}}\right>$ [$\mathrm{days}$] & $\left<\Delta S_{\mathrm{obs}}\right>$ [$\mathrm{Jy}$]  \\  
\hline 
4.8 & UMRAO & 46 & 0.09\\
8.0 & UMRAO & 38 & 0.07\\
14.5 &UMRAO & 32 & 0.07\\
37 & Mets\"ahovi & 16 & 0.24 \\
230 & SMA & 27 & 0.18 \\
340 & SMA & 132 & 0.25 \\
\hline
\end{tabular} 
\end{table} 

Figure~\ref{lightcurve} shows the total flux densities measured by the telescopes at different frequencies. The most prominent 
feature in the light curve is the major flare around 2006.2, best seen at 37\,$\mathrm{GHz}$. This feature is surrounded 
by smaller flares in 2005.2, 2007.6, 2008.5, and 2009.4. The light curves show the typical evolution of a flare: the flaring 
phenomenon usually starts at high frequencies and propagates to lower frequencies with a certain time delay of the peak, but there 
are also flares that develop simultaneously over a wide frequency range. The flare around 2005.0 appears nearly simultaneously at 
the highest frequencies (230\,$\mathrm{GHz}$ and 340\,$\mathrm{GHz}$) and delayed at 37\,$\mathrm{GHz}$ and 14.5\,$\mathrm{GHz}$, 
whereas it seems that the lowest frequencies (4.8\,$\mathrm{GHz}$ and 8\,$\mathrm{GHz}$) are not affected by this event.

The main flare shows the typical evolution: it is clearly visible at all 
frequencies with increasing time delays towards lower frequencies. Due to the poor time sampling of the 340\,$\mathrm{GHz}$ 
observations, the rise and the {time shift} at this frequency are not {easy to determine}. The remarkable double peak structure of 
the 230\,$\mathrm{GHz}$ measurements is an interesting feature and will be discussed later. After the flare, the flux decreases at all 
frequencies, with a steeper descent at higher frequencies.

\subsection{Time sampling and interpolation}
To perform a spectral analysis of the light curves, simultaneous data points are needed. This was achieved by performing a linear 
interpolation between the flux density values from the observations. The choice of an adequate time sampling, $\Delta t$, depends on 
the time interval of the observations and the significance of the frequency for the determination of the {peak frequency, i.e., the frequency where the spectral shape changes from optically thick to optically thin}. If a too short time interval is chosen, most of the data points would be interpolated ones. On the opposite 
situation, if a too long time interval is chosen, most of the light curves would be smoothed. Moreover, in this case, the data are
sampled inhomogeneously at the different frequency, so a compromise is required. The influence of a certain frequency on the 
calculation of $\nu_m$ can be estimated from the light curve (see Fig.~\ref{lightcurve}). The turnover 
frequency is usually between the frequencies of the highest and second highest flux density at a certain epoch. In the case studied
here, the dominating frequencies are 37\,$\mathrm{GHz}$ and 230\,$\mathrm{GHz}$. 
From this distribution it could be concluded that an appropriate time sampling should not be significantly shorter than the 
observational time interval used for these frequencies. A time sampling of $\Delta t=0.05\,\mathrm{yr}$ was selected for the 
interpolation, corresponding approximately to the observation cadence of the 37\,$\mathrm{GHz}$-light curve.

The correct handling of the uncertainties in the interpolated flux densities requires knowledge of a mathematical 
relation describing the light curves. Since this approach is out of the scope of this paper, we assigned 
the maximum uncertainty of the two closest observed flux densities to the interpolated flux density. {For our analysis we focused on the time interval between 2005.6 and 2006.8.} The interpolation of the poorly sampled $340\,\mathrm{GHz}$ light curve after $2006.2$ could induce artificially flat 
spectra. We took this fact into consideration by excluding the interpolated $340\,\mathrm{GHz}$ flux 
densities from the spectral analysis for $2006.2<t<2006.8$.

\subsection{Spectral analysis}
\label{specana}
Following  \citep{Turler:2000p1}, a synchrotron self absorbed spectrum is described by

\begin{equation}
S_\nu=S_m\left(\frac{\nu}{\nu_m}\right)^{\alpha_t}\frac{1-\exp{\left(-\tau_m\left(\nu/\nu_m\right)^{\alpha_0-\alpha_t}\right)}}{1-\exp{(-\tau_m)}},
\label{snu}
\end{equation}
where $\tau_m\approx3/2\left(\sqrt{1-\frac{8\alpha_0}{3\alpha_t}}-1\right)$ is the optical depth at the turnover 
frequency, $S_m$ is the turnover flux density, $\nu_m$ is the turnover frequency and $\alpha_t$ and 
$\alpha_0$ are the spectral indices for the optically thick and optically thin parts of the spectrum, respectively. 

The observed spectra {may in general be thought as the superposition of the emission from the steady state and perturbed (shocked) 
jet. We re-constructed this quiescent spectrum from archival data and we identified the quiet state with the minimum flux density of the low frequency light curves ($4.8\,\mathrm{GHz}$ - $37\,\mathrm{GHz}$) around $t=1989.0$ (see Fig. \ref{archival} and Table \ref{qstate}). The flux densities were fitted by a power law $S(\nu)=c_{q}\nu^{\alpha_{0}}$ and we obtained $c=7.43\pm0.65$ and $\alpha_{0}=-0.45\pm0.04$}.

\begin{table}[h!] 
\caption{Frequencies and flux density values for the quiescent spectrum}  
\label{qstate}
\centering  
\begin{tabular}{c c c c} 
\hline\hline
$\nu$ [$\mathrm{GHz}$]& $S$ [$\mathrm{Jy}$] & Observatory\\  
\hline 
4.8	& 3.68$\pm$0.05	& UMRAO\\
8.0	& 2.95$\pm$0.06	&UMRAO \\
14.5	& 2.10$\pm$0.06	&UMRAO\\
22	& 2.12$\pm$0.14	& Mets\"ahovi\\
37	& 1.62$\pm$0.11	& Mets\"ahovi\\
\hline
\end{tabular} 
\end{table}

\begin{figure}[h!]
\resizebox{\hsize}{!}{\includegraphics{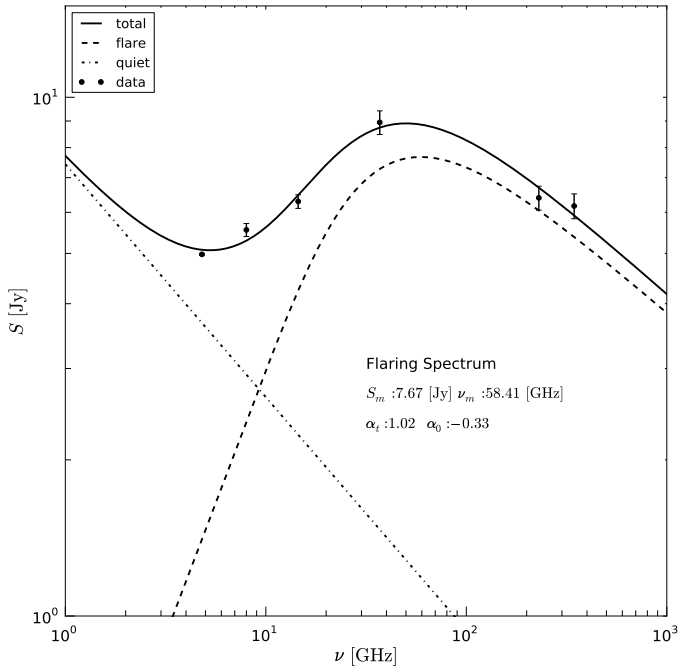}} 
\caption{Result of the spectral fitting to the 2006.20 data. The dashed-dotted line corresponds to the quiescent spectrum, the 
dashed line to the flaring spectrum, and the solid black line to the total spectrum. The values presented indicate the spectral 
turnover of the flaring spectrum.} 
\label{2006.35} 
\end{figure}

For the spectral analysis, we removed the contribution of the quiescent spectrum from the interpolated data points 
and applied Eq. ~\ref{snu}. The uncertainties of the remaining flaring spectrum were calculated using the errors of 
the interpolated data points and the obtained uncertainties of the quiescent spectrum. During the fitting process 
we allowed both spectral indices, $\alpha_t$ and $\alpha_0$ to vary. Figure~\ref{2006.35} shows the result 
of a spectral fitting at a selected epoch applied to interpolated light curve data.

\subsection{Error analysis of the spectra}
Due to the non-linear nature of Eq.~\ref{snu} we applied a Monte-Carlo simulation to the observed/interpolated flux 
densities in order to derive estimates for the uncertainties of the fitting parameters ($S_m$, $\nu_m$, $\alpha_t$, 
and $\alpha_0$). Random values for each simulated spectrum were drawn from a Gaussian distribution with a mean, 
$\mu$, equal to the observed flux density, $S(\nu)$, and a variance, $\sigma$, equal to the observed flux density 
error, $\Delta S(\nu)$. In order to achieve a good statistical ensemble, up to 1000 spectra were simulated per 
interpolated spectrum. Each of these simulated spectra were fitted with a synchrotron spectrum (see Eq. \ref{snu}). 
As the expected value and uncertainty of a spectral parameter we take the
mean and standard deviation of its simulated probability density
distribution, respectively. The derived spectral parameters and their
uncertainties are presented in Table \ref{finaldata}.

\begin{table*}
\caption{Spectral parameters (expected values) and corresponding uncertainties from the Monte-Carlo simulation}
\label{finaldata}
\centering
\begin{tabular}{c c c c c}
\hline\hline
$t$ [yr] & $\nu_m$ [GHz] & $S_m$ [Jy] & $\alpha_t$ & $\alpha_0$\\
\hline 
2005.60 & 222$\pm$99 &  3.3$\pm$0.3 &  0.33$\pm$0.14 & $-$0.27$\pm$0.50 \\ 
2005.65 & 163$\pm$27 &  5.9$\pm$0.5 &  0.56$\pm$0.11 & $-$1.80$\pm$0.57 \\ 
2005.70 & 135$\pm$25 &  5.7$\pm$0.8 &  0.59$\pm$0.15 & $-$0.90$\pm$0.50 \\ 
2005.75 & 119$\pm$19 &  6.5$\pm$1.1 &  0.70$\pm$-0.16 & $-$0.78$\pm$0.42 \\ 
2005.80 & 109$\pm$12 &  8.4$\pm$1.0 &  0.78$\pm$0.14 & $-$1.02$\pm$0.28 \\ 
2005.85 & 96$\pm$12 &  7.7$\pm$1.0 &  0.82$\pm$0.16 & $-$0.72$\pm$0.25 \\ 
2005.90 & 88$\pm$14 &  6.7$\pm$0.9 &  0.81$\pm$0.16 & $-$0.42$\pm$0.28 \\ 
2005.95 & 73$\pm$14 &  6.1$\pm$0.5 &  1.0$\pm$0.3 & $-$0.22$\pm$0.15 \\ 
2006.00 & 62$\pm$12 &  6.2$\pm$0.5 &  1.2$\pm$0.3 & $-$0.20$\pm$0.11 \\ 
2006.05 & 56$\pm$11 &  6.5$\pm$0.6 &  1.3$\pm$0.4 & $-$0.19$\pm$0.11 \\ 
2006.10 & 63$\pm$12 &  6.5$\pm$0.6 &  1.1$\pm$0.3 & $-$0.20$\pm$0.13 \\ 
2006.15 & 60$\pm$9 &  7.2$\pm$0.6 &  1.0$\pm$0.3 & $-$0.28$\pm$0.13 \\ 
2006.20 & 57$\pm$7 &  7.9$\pm$0.6 &  1.1$\pm$0.2 & $-$0.33$\pm$0.12 \\ 
2006.25 & 56$\pm$7&  8.5$\pm$0.8 &  0.93$\pm$0.22 & $-$0.47$\pm$0.20 \\ 
2006.30 & 56$\pm$7 &  8.4$\pm$0.7 &  0.94$\pm$0.21 & $-$0.43$\pm$0.19 \\ 
2006.35 & 53$\pm$7 &  8.4$\pm$0.7 &  0.97$\pm$0.24 & $-$0.45$\pm$0.20 \\ 
2006.40 & 41$\pm$5 &  7.2$\pm$0.6 &  1.0$\pm$0.3 & $-$0.49$\pm$0.13 \\ 
2006.45 & 27$\pm$6 &  4.5$\pm$0.3 &  1.1$\pm$0.5 & $-$0.35$\pm$0.13 \\ 
2006.50 & 25$\pm$7 &  3.8$\pm$0.4 &  0.89$\pm$0.56 & $-$0.40$\pm$0.18 \\ 
2006.55 & 25$\pm$7 &  3.4$\pm$0.3 &  0.37$\pm$0.38 & $-$0.66$\pm$0.32 \\ 
2006.60 & 25$\pm$10 &  2.9$\pm$0.4 &  0.28$\pm$0.35 & $-$0.61$\pm$0.35 \\ 
2006.65 & 25$\pm$9 &  2.6$\pm$0.4 &  0.24$\pm$0.19 & $-$0.47$\pm$0.34 \\ 
2006.70 & 21$\pm$11 &  2.3$\pm$0.4 &  0.26$\pm$0.13 & $-$0.59$\pm$0.34 \\ 
2006.75 & 17$\pm$7 &  2.0$\pm$0.3 &  0.22$\pm$0.18 & $-$0.63$\pm$0.33 \\ 
2006.80 & 16$\pm$8 &  1.8$\pm$0.3 &  0.24$\pm$0.11 & $-$0.75$\pm$0.43 \\ 
\hline
\end{tabular}
\end{table*}

\section{Synchrotron radiation and the shock-in-jet model}
\label{theory}
In this section we review the shock-in-jet model as derived by \citet{Marscher:1985p50} and different modifications
to it. In this review we have included the redshift dependence in the different equations where it is required. In 
this way, we can fit the model to observational data from any source. 

\subsection{Synchrotron radiation}
The specific intensity from a power law distribution of relativistic electrons, $N(E)=KE^{-s}$, where $K$ is the 
normalization coefficient of the distribution and $s$ the spectral slope is defined by:
\begin{equation}
I_\nu=\frac{\epsilon_\nu}{\kappa_\nu}\left(1-e^{-\tau_{\nu}}\right),
\end{equation}
where $\epsilon_\nu$ and $\kappa_\nu$ are the emission and absorption coefficients, respectively, and $\tau_\nu$ 
is the optical depth. The emission and absorption coefficients can be written as 
\citep[for details see][]{Pacholczyk:1970p1583}:
\begin{eqnarray}
\epsilon_\nu&\propto&K B^{(s+1)/2}\nu^{-(s-1)/2}\\
\kappa_\nu&\propto&K B^{(s+2)/2} \nu^{-(s+4)/2},
\end{eqnarray}
where $B$ is the magnetic field and $\nu$ the frequency in the source frame. With this definition of the emission and 
absorption coefficients the specific intensity is:
\begin{equation}
I_{\nu}\propto B^{-1/2}\nu^{5/2}\left(1-e^{-\tau_{\nu}}\right).
\label{approx1}
\end{equation}
Depending on the optical depth, $\tau_{\nu}$, Eq. \ref{approx1} describes an optically thick $(\tau_{\nu}>1)$ or 
optically thin $(\tau_{\nu}<1)$ spectrum with their characteristic shapes  $I_{\nu}\propto \nu^{5/2}$ and  
$I_{\nu}\propto \nu^{-(s-1)/2}$, respectively.\\

The optically thin flux density from a jet located at luminosity distance $D_{L}$, with radius $R$, conical geometry 
(half opening angle $\phi$), constant velocity $\beta$ and viewing angle $\vartheta$ is given by
\begin{equation}
S_\nu=\Omega x \epsilon_\nu,
\label{snu1}
\end{equation}
where $\Omega=\pi R^2/D_L^2(1+z)^{4}$ is the solid angle, and $z$ was the redshift. Taking into account the influence 
of the Doppler factor, $D=\Gamma^{-1}\left(1-\beta\cos\vartheta\right)^{-1}$ \citep[see e.g.,][]{Lind:1985p3025}, where 
$\Gamma=\left(1-\beta^{2}\right)^{-1/2}$ the Lorentz factor, Eq. \ref{snu1} leads to:
\begin{equation}
S_\nu\propto (1+z)^{-(s-3)/2}D_L^{-2}R^2xKB^{(s+1)/2}D^{(s+3)/2}\nu^{-(s-1)/2}.
\label{su}
\end{equation}
The turnover frequency, defined as the frequency where the optical depth is equal to unity, can be expressed as: 
\begin{equation}
\nu_m \propto(1+z)^{-1}\left(xKB^{(s+2)/2}D^{(s+2)/2}\right)^{2/(s+4)}.
\label{nu1}
\end{equation}
The turnover flux density $S_{m}\left(\nu_{m}\right)$ is then obtained by inserting Eq. \ref{nu1} into Eq. \ref{su}:
\begin{equation}
S_{m}\propto (1+z)D_{L}^{2}R^{2}\left(K^{5}B^{2s+3}x^{5}D^{3s+7}\right)^{1/(s+4)}.
\label{smn}
\end{equation}
The equations above were successfully used to model the steady-state emission from quasars, where a decrease in the 
particle density $N$ and the magnetic field $B$ along the jet was assumed \citep[e.g.,][]{Konigl:1981p15}. However, 
the observed flux densities during flares showed a more complicated behavior and could not be described by 
steady-state models.

\subsection{Shock-in-jet model}
\label{shockinjet}
The shock-in-jet model of \citet{Marscher:1985p50} studied the evolution of a traveling shock wave in a steady state 
jet. During the passage of the shock through a steady jet, the relativistic particles are swept up at the 
shock front and gain energy while crossing it. In this model, the flaring flux density is assumed to be produced by the accelerated 
particles within a small layer of width $x$ behind the shock front. The width of this layer is assumed to depend on 
the dominant cooling process and can be approximated by $x\propto t_{\mathrm{cool}}$, where $t_{\mathrm{cool}}$ is 
the typical cooling time i.e., synchrotron cooling time.\\

\subsubsection{Evolutionary stages}
\paragraph{Compton losses:}
Assuming the photon energy density, $u_{\mathrm{ph}}$, is higher than the magnetic energy density, $u_{\mathrm{b}}=B^{2}/(8\pi)$, the inverse Compton scattering is the dominant energy loss mechanism during the first 
stage of the flare. The width of the layer behind the shock front during this so-called Compton stage, $x_{1}$, is 
computed to be
\begin{equation}
x_{1}\propto B^{1/2}\nu^{-1/2} D^{1/2}u_{\mathrm{ph}}^{-1}.
\label{x1n}
\end{equation}
An approximation of the photon energy density, $u_{\mathrm{ph}}$, can be obtained by integrating the emission 
coefficient, $\epsilon_{\nu}$, over the optically thin regime 
($\nu_{m}<\nu<\nu_{\mathrm{max}}$)\footnote{ \citet[][]{Marscher:1985p50} included only first order Compton scattering}, 
which leads to
\begin{equation}
u_{\mathrm{ph}}\propto K\left(B^{3s+7}R^{s+5}\right)^{1/8}.
\end{equation}
The final expression for the width of the layer behind the shock front during the Compton stage can be written as
\begin{equation}
x_1\propto R^{-(s+5)/8} K^{-1}B^{-3(s+1)/8}D^{1/2}\nu^{-1/2}.
\end{equation}

\paragraph{Synchrotron losses:}
Synchrotron losses become more important at the point where the photon energy density, $u_{\mathrm{ph}}$, is 
comparable to the magnetic energy density, $u_{\mathrm{b}}$. The width of the layer $x_{2}$ in the synchrotron stage 
can be computed by replacing $u_{\mathrm{ph}}$ in Eq. \ref{x1n} with  $u_{\mathrm{b}}=B^{2}/(8\pi)$ and is given by
\begin{equation}
x_2\propto B^{-3/2}D^{1/2}\nu^{-1/2}.
\end{equation}

\paragraph{Adiabatic losses:}
Radiative losses become less important in the last stage of the shock evolution. During this final stage the losses 
are dominated by the expansion of the source and the width of the layer, $x_{3}$, is assumed to be similar to the 
radius of the jet
\begin{equation}
x_{3}\propto R.
\end{equation}
The evolution of the turnover frequencies $\nu_{m,i}$ and turnover flux densities $S_{m,i}$, where $i$ indicates the 
different energy loss stages (1=Compton, 2=synchrotron and 3=adiabatic loss stage), can be derived by inserting the 
expressions for $x_{i}$ into Eq. \ref{nu1} and Eq. \ref{smn}:
\begin{eqnarray}
\nu_{m,1}&\propto& \left(1+z\right)^{-(s+4)/(s+5)} R^{-1/4}B^{1/4}D^{(s+3)/(s+5)}\\
S_{m,1}&\propto&\left(1+z\right)^{(2s+15)/(2s+10)} D_L^{-2} R^{11/8}B^{1/8}D^{(3s+10)/(s+5)}\\
\nu_{m,2}&\propto& \left(1+z\right)^{-(s+4)/(s+5)} \left[K^2B^{s-1}D^{s+3}\right]^{1/(s+5)}\\
S_{m,2}&\propto&\left(1+z\right)^{(2s+15)/(2s+10)} D_L^{-2} R^{2}\left[K^5B^{2s-5}D^{3s+10}\right]^{1/(s+5)}\\
\nu_{m,3}&\propto& \left(1+z\right)^{-1} \left[RKB^{(s+2)/2}D^{(s+2)/2}\right]^{2/(s+4)}\\
S_{m,3}&\propto& \left(1+z\right)D_L^{-2} \left[R^{2s+13}K^5B^{2s+3}D^{3s+7}\right]^{1/(s+4)}
\end{eqnarray}
Furthermore \citet{Marscher:1990p3333,Lobanov:1999p2299} assumed that the evolution of $K$, $B$ and $D$ could be written as a power-law 
with the jet radius $R$ (notice that in a conical jet the distance along the jet $L$ is linearly proportional to the 
jet radius $R$, $L\propto R$),
\begin{equation}
K\propto R^{-k} \quad B\propto R^{-b} \quad D\propto R^{-d}\, .
\label{prop}
\end{equation} 
Then, a relation between the turnover flux density and the turnover frequency can be found:
\begin{eqnarray}
\nu_{m,i}&\propto& (1+z)^{p_i} R^{n_i}
\label{numr}\\
S_{m,i}&\propto& D_L^{-2}(1+z)^{q_i} R^{f_i},
\label{smr}
\end{eqnarray}
which leads to 
\begin{equation}
S_{m,i}\propto D_L^{-2}\left(1+z\right)^{(q_in_i-p_if_i)/n_i}\nu_{m,i}^{\epsilon_{i}},
\label{smnum}
\end{equation}
where $\epsilon_{i}={f_i}/{n_i}$.
The exponents $f_i$, $n_i$, $p_i$ and $q_i$ include the dependence with the physical quantities $B$, $N(E)$, $K$, and 
$D$ and can be written as:
\begin{eqnarray}
n_1&=&-(b+1)/4-d(s+3)/(s+5) \label{an1}\\
n_2&=&-[2k+b(s-1)+d(s+3)]/(s+5)\\
n_3&=&-[2(k-1)+(b+d)(s+2)]/(s+4)\label{n3}\\
f_1&=&(11-b)/8-d(3s+10)/(s+5)\\
f_2&=&2-[5k+b(2s-5)+d(3s+10)]/(s+5)\\
f_3&=&[2s+13-5k-b(2s+3)-d(3s+7)]/(s+4)\label{f3} \\
p_1&=&-(s+4)/(s+5)\\
p_2&=&-(s+4)/(s+5)\\
p_3&=&-1\\
q_1&=&(2s+15)/(2s+10)\\
q_2&=&(2s+15)/(2s+10)\\
q_3&=&1 \label{aq3}.
\end{eqnarray}
The typical evolution of a flare in the turnover frequency -- turnover flux density $(\nu_{m}-S_{m})$ plane can be 
obtained by inspecting the $R$-dependence of the turnover frequency, $\nu_{m}$, and the turnover flux density, 
$S_{m}$, for a typical set of parameters: $b$=1, $s=2.5$, $k=3$ (assuming an adiabatic flow, $k=2(s+2)/3$) and $d=0$ 
(assuming constant velocity). These parameters lead to a set of exponents $n_{i}$ and $f_{i}$, which are summarized 
in Table~\ref{nifi}.

\begin{table}[h!]
\caption{Values of the indices used in the generalized spectrum expression of Eq. \ref{smnum}, $n_{i}$ and $f_{i}$ 
for $b=1$, $s=2.5$, $k=3$, and $d=0$}
\label{nifi}
\begin{tabular}{l c c c}
\hline \hline
$stage$ & $n_{i}$ & $f_{i}$ & $\epsilon$ \\
\hline
\noalign{\smallskip}
Compton & $-$0.5 & 1.3 & $-$2.6\\
Synchrotron & $-$1 & 0 & 0\\
Adiabatic & $-$1.3 & $-$0.8 & 0.6\\
\hline
\end{tabular}
\end{table}

During the first stage, where Compton losses are dominant, the turnover frequency, $\nu_{m}$, decreases with 
radius, $R$, while the turnover flux density, $S_{m}$, increases. In the second stage, where synchrotron losses are 
the dominating energy loss mechanism, the turnover frequency continues to decrease while the turnover 
flux density is constant. Both the turnover frequency and turnover flux density decrease in the final, adiabatic stage.

\subsubsection{Generalization of the shock-in-jet model}
\citet{Turler:2000p1} expanded the shock-in-jet model to non-conical jets, e.g., the distance along the jet, $L$, 
is no longer a linearly proportional to the jet radius, $R$. This modification is expressed by 
$R\propto L^{r}$ with ($-1<r<1$), which represents a collimating ($r<0$) or expanding ($r>0$) jet.
An important implication of allowing non-conical jets ($r\neq1$) in the shock-in-jet model is the $r$-dependence of 
the evolution of the spectral parameters ($b$, $k$ and $d$). Therefore the evolution of $B$, 
$K$ and $D$ along the jet is given by:

\begin{equation}
K\propto L^{-rk}, \quad B\propto L^{-rb}, \quad D\propto L^{-rd}.
\end{equation} 

Besides the modification of the jet geometry, \citet{Turler:2000p1} parametrized the temporal evolution of the 
turnover frequency, $\nu_{m}$, and the turnover flux density, $S_{m}$, using the equations of superluminal 
motion\footnote{The time used here was that in the observers frame.}
\begin{equation}
t=\frac{L}{\mathrm{v}_{\mathrm{obs}}}\, \mathrm{ where }\, \mathrm{v}_\mathrm{obs}=\frac{\beta_\mathrm{app}c}{(1+z)\sin\vartheta}. 
\end{equation}
Using the definition of the apparent speed, $\beta_\mathrm{app}=(\beta\sin\vartheta)/(1-\beta\cos\vartheta)$, the bulk Lorentz 
factor, $\Gamma$, and the Doppler factor, $D$, the equation above results in
 \begin{equation} 
 t=\frac{1+z}{\beta cD\Gamma}L.
 \label{t1}
 \end{equation}
Assuming that $\beta\sim1$ and $\vartheta\sim 1/\Gamma \ll1$, it can be shown that $\Gamma\propto D$ 
(Taylor expanding $\cos{1/\Gamma}$ in the definition of $D$). Including the non-conical jet geometry 
($R\propto L^{r}$) the equation above can be written as
\begin{equation}
t\propto(1+z)D^{-2}R^{1/r}\propto (1+z)R^\rho,
\label{time}
\end{equation}
where $\rho=(2dr+1)/r$. By replacing $R$ in Eqs.~\ref{numr} and \ref{smr} by Eq.~\ref{time}, the temporal evolution of the 
turnover frequency and turnover flux density is:
\begin{eqnarray}
\nu_m&\propto& (1+z)^{(p_i\rho-n_i)/\rho}\cdot t^{n_i/\rho} \label{tnum}\\
S_m&\propto& D_L^{-2}(1+z)^{(q_i\rho-f_i)/\rho}\cdot t^{f_i/\rho}.
\end{eqnarray}
The equations above can be further simplified by replacing $D_L\propto (1+z)$, which leads to:
\begin{equation}
S_m\propto (1+z)^{\left[(q_i-2)\rho-f_i\right]/\rho}\cdot t^{f_i/\rho}.
\label{tsm}
\end{equation}

\subsubsection{Modification of the compton stage}
The determination of the distance that the relativistic particles travel behind the shock front before 
losing most of their energy, was a crucial parameter in shock-in-jet models.
\citet{Bjornsson:2000p32} presented a more accurate derivation of this distance than previous works, 
including the possibility of multiple Compton scattering (in the Thompson regime) during the first rising 
phase of the flare. In the case of first order Compton scattering, the evolution of the first 
stage was predicted to be less abrupt when compared to \citet{Marscher:1985p50}. Therefore 
the existence of a synchrotron stage was no longer needed. The evolution of the turnover-flux 
density, $\nu_{m}$, and the turnover frequency, $S_{m}$, with distance (or time) is given, within this model, by:

\begin{eqnarray}
\nu_m&\propto& R^{-[4(s+2)+3b(s+4)]/[3(s+12)]}
\label{numbj}\\
S_m&\propto& R^{-[4(s-13)+6b(s+2)]/[3(s+12)]},
\label{smbj}
\end{eqnarray}
assuming no changes in the Doppler factor and constant velocity of the shocked particles. Using the same set of 
parameters as in Table \ref{nifi} leads to $n_{1}= -0.9$ and $f_{1}=0.3$, resulting in a less steep slope 
$\epsilon_{1}=f_{1}/n_{1}=-0.3$ in the $\nu_{m}-S_{m}$-plane which is flatter than in \citet{Marscher:1985p50} case.

\section{Results}
\label{results}
\begin{figure}[h!]
\resizebox{\hsize}{!}{\includegraphics{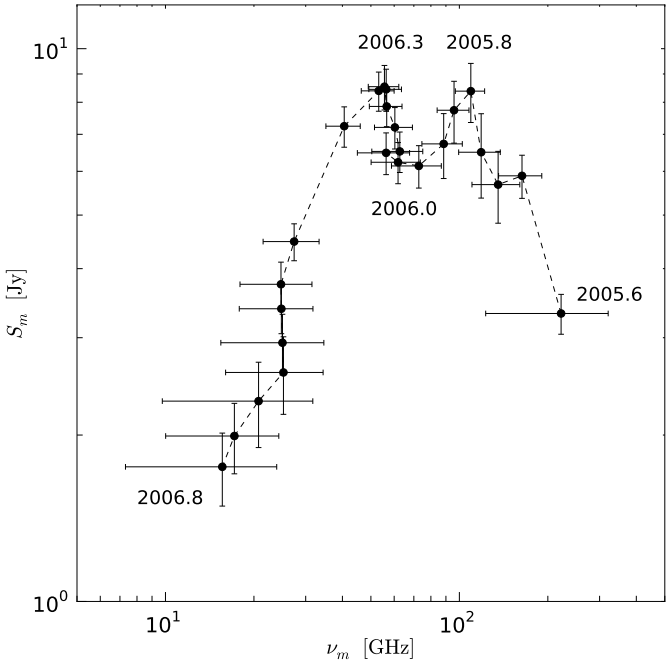}} 
\caption{The 2006 radio flare in the turnover frequency - turnover flux density plane. The time labels indicate the time evolution 
and the temporal position of local and global extrema in the spectral evolution.} 
\label{numsm} 
\end{figure}

\begin{figure}[h!]
\resizebox{\hsize}{!}{\includegraphics{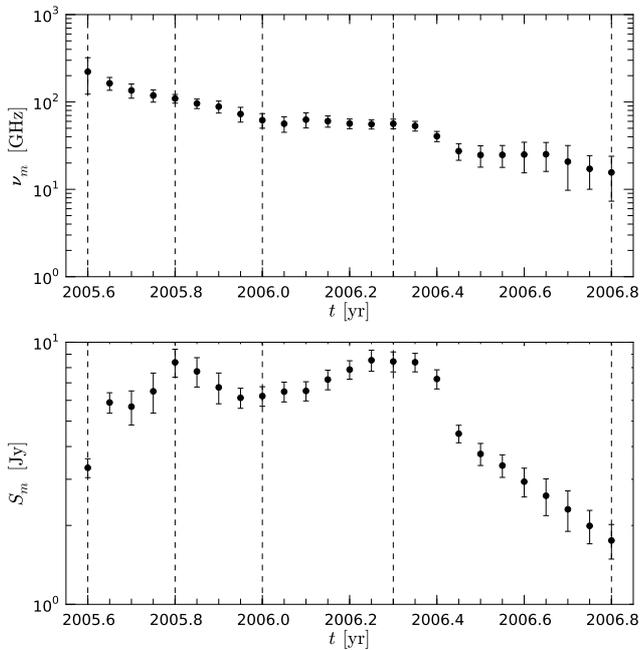}} 
\caption{Temporal evolution of the 2006 radio flare: top turnover frequency and bottom turnover flux density. The dashed vertical lines correspond to the time labels in Figure \ref{numsm} and indicate the extrema in the evolution.} 
\label{numsmtime} 
\end{figure}

\begin{figure}[h!]
\resizebox{\hsize}{!}{\includegraphics{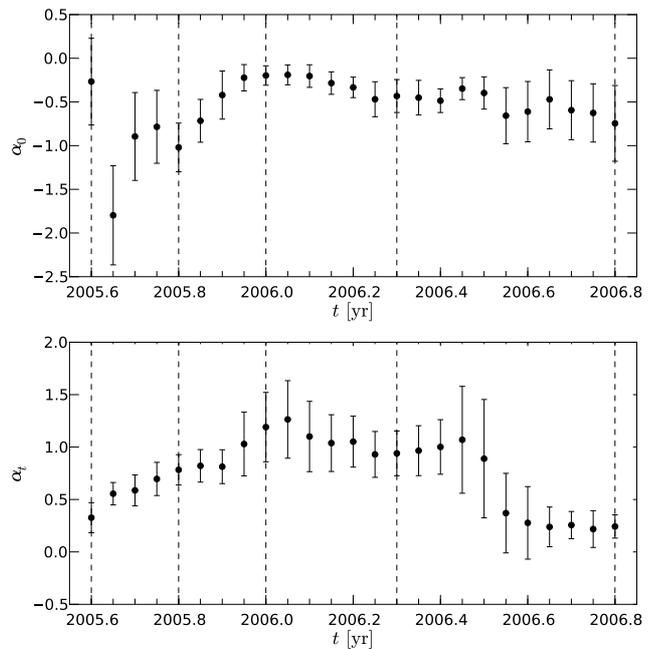}} 
\caption{The temporal evolution of the optically thin ($\alpha_0$, top) and thick ($\alpha_t$, bottom) spectral indices for 2006 radio flare. The dashed vertical lines corresponds to the time labels in Figure \ref{numsm} and indicate the extrema in the evolution.} 
\label{alpha} 
\end{figure}

The spectral evolution of the 2006 radio flare in CTA\,102 is presented in the turnover frequency - turnover flux
 density ($\nu_m-S_m$) plane, shown in Fig.~\ref{numsm}. Figures~\ref{numsmtime} and \ref{alpha} display the time 
evolution of the spectral parameters $\left(\nu_m, S_m, \alpha_0,\, \mathrm{ and }\, \alpha_t \right)$.

In the next paragraphs, we summarize the evolution of the event in the turnover frequency --- turnover flux density 
plane and compare it with the standard shock-in-jet model \citep{Marscher:1985p50} and later modifications to it
\citep{Bjornsson:2000p32}. This information is also summarized in Table~\ref{section}.

The flare started around 2005.6 with a high turnover frequency ($\nu_m\sim 200\,\mathrm{GHz}$) and a low turnover 
flux density ($S_m\sim 3\,\mathrm{Jy}$). During the first {0.2\,$\mathrm{yr}$}, the turnover flux density, $S_m$, 
increased, reaching $S_m\sim 8.4\,\mathrm{Jy}$, while the turnover frequency decreased to 
$\nu_m\sim 110\,\mathrm{GHz}$. The slope of the optically thick part of the spectrum, represented by $\alpha_{t}$, 
steepened (0.33 to 0.78), while the optically thin spectral index, $\alpha_{0}$, {steepened during the first $0.05\,\mathrm{yr}$ and flattened ($-$1.8 to $-$1.02) afterwards}. The large uncertainties 
of the optically thin spectral index, $\alpha_{0}$, are due to the lack of data points beyond 340\,$\mathrm{GHz}$.
{The exponential relation between the turnover flux density, $S_{m}$ and the turnover frequency, $\nu_{m}$, 
($S_{m}\propto \nu_{m}^{\epsilon_{i}}$, see Sect. \ref{shockinjet}), led to a value of $\epsilon=-1.21\pm0.22$.}

{As mentioned in Sect. \ref{theory}, the dominant loss mechanism during the first stage of flare (increasing turnover flux density while the turnover frequency was decreasing) was Compton scattering.}\citet{Marscher:1985p50} predicted a value of $\epsilon=-5/2$, whereas \citet{Bjornsson:2000p32} derived 
$\epsilon=-0.43$ using a modified expression for the shock width (both assumed $s=2.4$ and $b=1$). The obtained 
value of $\epsilon=-1.21\pm0.22$ was in between these two values, but it is impossible to reproduce using the 
approach of \citet{Bjornsson:2000p32}. 

During the time interval between 2005.8 and 2006.0, involving 0.2\,$\mathrm{yr}$, the turnover flux density 
and turnover frequency decreased to {$S_m\sim 6.2\,\mathrm{Jy}$ and to $\nu_m\sim 62\,\mathrm{GHz}$}, respectively. 
The slope in the $\nu_{m}-S_{m}$-plane changed to $\epsilon=0.77\pm0.11$. {The average optically thick spectral index reached 
a value $\alpha_{t}\sim 0.82\pm0.17$ while the optically thin spectral index, $\alpha_{0}$, continued rising
to $-$0.22.} This behavior of the turnover values fits well in 
the adiabatic stage in the shock-in-jet model. For this stage \citet{Marscher:1985p50} derived an exponent 
$\epsilon=0.69$ (assuming $s=3$), which is well within our value.

However, the increase of the turnover flux density starting in 2006.0, which reached a peak value of 
$S_m\sim 8.5\,\mathrm{Jy}$ in 2006.3, cannot be explained within the frame of the shock-in-jet model. 
Its behavior, though, resembles that expected from a Compton stage ($\epsilon=-0.99\pm0.46$). 
The spectral indices during this stage were roughly constant, 
$\alpha_{t}=1.33\pm0.27$ and $\alpha_{0}=-0.26\pm0.13$.

After 2006.3, the turnover flux density and turnover frequency decreased. This last phase could be understood as an 
adiabatic stage and the power law fit in the $\nu_{m}-S_{m}$-plane led to an $\epsilon=1.24\pm0.10$. The optically 
thin spectral index was nearly constant, $\alpha_{0}=-0.46\pm0.21$ while the evolution of the optically thick 
spectral index, $\alpha_{t}$, could be divided into two parts: $\alpha_{t}=0.96\pm0.28$ until 2006.5 and 
$\alpha_{t}=0.26\pm0.18$ after.

Our results show no evidence for a synchrotron stage, characterized by a nearly constant turnover flux density, 
$S_{m}$, while the turnover frequency, $\nu_{m}$, is decreasing 
\citep[$\epsilon=-0.05$ for $s=2.4$ and $b=1$,][]{Marscher:1985p50}.

{The first hump in the evolution of the flare in the $\nu_{m}-S_{m}$-plane ($2005.6<t<2006.0$) fulfills, despite no evidence for a synchrotron stage, the phenomenological requirements of the shock-in-jet model of \citet{Marscher:1985p50}. However, the evolution after 2006.0 did not follow the predicted evolution. The turnover flux density and the turnover frequency should continue decreasing. But their behavior mimicked the evolution of a Compton stage ($2006.0<t<2006.3$) and the one of an adiabatic stage ($2006.3<t<2006.8$). Considering these points, we decided to model the evolution
of the 2006 flare in CTA\,102 according to the standard shock-in-jet model \citep{Marscher:1985p50} including a second Compton and adiabatic stage. This decision is evaluated in Sect.~\ref{disc}.}

\begin{table*}
\begin{minipage}[t]{\columnwidth}
\caption{Different stages of the spectral evolution and their characteristics}
\label{section}
\centering
\begin{tabular}{c c c c c c}
\hline\hline
time & stage &model name&$\epsilon$ $\left(S_{m}\propto\nu_{m}^{\epsilon}\right)$ & $\left<\alpha_t\right>$ &$\left<\alpha_0\right>$ \\
\hline
2005.6 -- 2005.8 & Compton& C1& $-$1.21$\pm$0.22 & rising 0.33 -- 0.78  & rising $-$1.80 -- $-$1.02\\
2005.8 -- 2006.0 & Adiabatic& A1& 0.77$\pm$0.11 & 0.82$\pm$0.17 & rising $-$1.02 -- $-$0.21 \\
2006.0 -- 2006.3 & Compton-like& C2& $-$0.99$\pm$0.46 & 1.03$\pm$0.27 & $-$0.26$\pm$0.13 \\
2006.3 -- 2006.8 & Adiabatic& A2& 1.24$\pm$0.10 & 0.96$\pm$0.29$^{a}$, 0.26$\pm$0.18$^{b}$ & $-$0.46$\pm$0.21\\
\hline
\end{tabular}
\begin{list}{}{}
\item[$^a$] until 2006.4, $^{b}$ after 2006.45
\end{list}
\end{minipage}
\end{table*}

\section{Modeling the 2006 radio flare}
\label{appmod}

In this section we present the results of applying the shock-in-jet model and the fitting technique to the observed spectral 
evolution of the 2006 radio flare in CTA\,102. 

\subsection{Fitting technique}
We used a multi-dimensional $\chi^2$-optimization for deriving a set of parameters that fit the spectral evolution of the 
different stages. Our approach consists on fitting the temporal evolution of the spectral turnover values, $\nu_m$ and $S_m$ using 
Eqs.~\ref{tnum} and \ref{tsm} and the definition of the spectral exponents (see Eqs.~\ref{an1}-\ref{aq3}). The proportionalities in 
Eqs.~\ref{tnum} and \ref{tsm} can be removed by introducing the constants $c_{\nu_{m,i}}$ and $c_{S_{m,i}}$, which reflect logarithmic 
shifts of the turnover frequency and flux density and depend on the intrinsic properties of the source and flare, having no further 
importance for our study. From the observed values, $\nu^{obs,j}_{m,i}$ and $S^{obs,j}_{m,i}$ ($i$ indicating the 
radiation loss stage, and $j$ indicating the position among the total number of points in the stage, $q$, so 
that $j=1\cdots q$), the proportionality constants can be derived as:

\begin{eqnarray}
c_{\nu_{m,i}}&=&\nu^{obs,j}_{m,i}\cdot t_{obs,j}^{-n_i/\rho},
\label{cnum}\\
c_{S_{m,i}}&=&S^{obs,j}_{m,i}\cdot t_{obs,j}^{-f_i/\rho},
\label{csm}
\end{eqnarray}
where $t_{obs,j}$ is the time in the observers frame.
Using these definitions, the constant for the spectral evolution in the $\nu_m-S_m$ plane (see Eq. \ref{smnum}) yields:

\begin{equation}
c_{\left(\nu_m-S_m\right),i}=S^{obs,j}_{m,i}\cdot \left(\nu^{obs,j}_{m,i}\right)^{-f_i/n_i}\cdot t_{obs,j}.
\end{equation}

From Eqs.~\ref{an1}-\ref{aq3}, \ref{tnum}, and \ref{tsm}, we see that there are 5 parameters 
($b$, $s$, $k$, $d$, and $r$) that describe the whole spectral evolution, although they have different values at each stage. 
Starting from basic physical principles, we used boundaries for the different parameters to avoid unphysical results. These boundaries
are listed in Table~\ref{boundaries}.

\begin{table}[h!]
\caption{Range for spectral parameters allowed to vary in the fits to the observed spectrum for all stages.}
\label{boundaries}
\begin{tabular}{c c c c c}
\hline \hline
$b$ & $s$ & $k$ & $d$ & $r$\\
\hline
\noalign{\smallskip}
1 to 2$^a$ & 2 to 4& 1 to 6& $-$2 to 2 & -1 to 1\\
\hline
\end{tabular}
\begin{list}{}{}
\item[$^a$] if $r>0$.
\end{list}
\end{table}

The negative values for parameter $d$ stand for the possibility for an increase in the Doppler 
factor, $D$, with radius. Regarding $r$, negative values correspond to collimation, i.e., a decrease of the jet 
radius, and positive values correspond to an expansion process. 

On top of the listed limitations, the evolution of the optically thin spectral index, $\alpha_0$,  can be used to 
provide estimates of the parameter $s=1-2\alpha_0$. \citet{Marscher:1985p50} used a lower limit of $s=2$ to keep 
the shock non-radiative. A flatter spectral slope, e.g., $s<2$, would increase the amount of high energy electrons, 
which will dominate the energy density and the pressure in shock. Since these particles suffer radiative cooling, 
their energy losses would affect the dynamics leading to a radiative shock.

As mentioned before, allowing non-conical jet expansion ($r\neq1$) also affects the boundaries for the spectral 
parameters $k$ and $d$ presented in Table~\ref{boundaries}, which depend on the value of $r$ in the non-conical case. Note that this should be
taken into account in order to properly interpret the evolution of the physical parameters with distance to the 
core.

We developed a least-square algorithm to fit the three different radiation loss stages (Compton, synchrotron, or adiabatic, 
or combinations of them) to the observed evolution using one single set of parameters. During the optimization of 
$\chi^2$ all observed data points were used for calculating the constants $c_{\nu_{m,i}}$ and $c_{S_{m,i}}$, which allows for an 
improved $\chi^2$. This technique allowed us to test and analyze different possible scenarios ($b$, $s$, $k$, $d$, and $r$). 

\begin{figure*}  
\includegraphics[width=17cm]{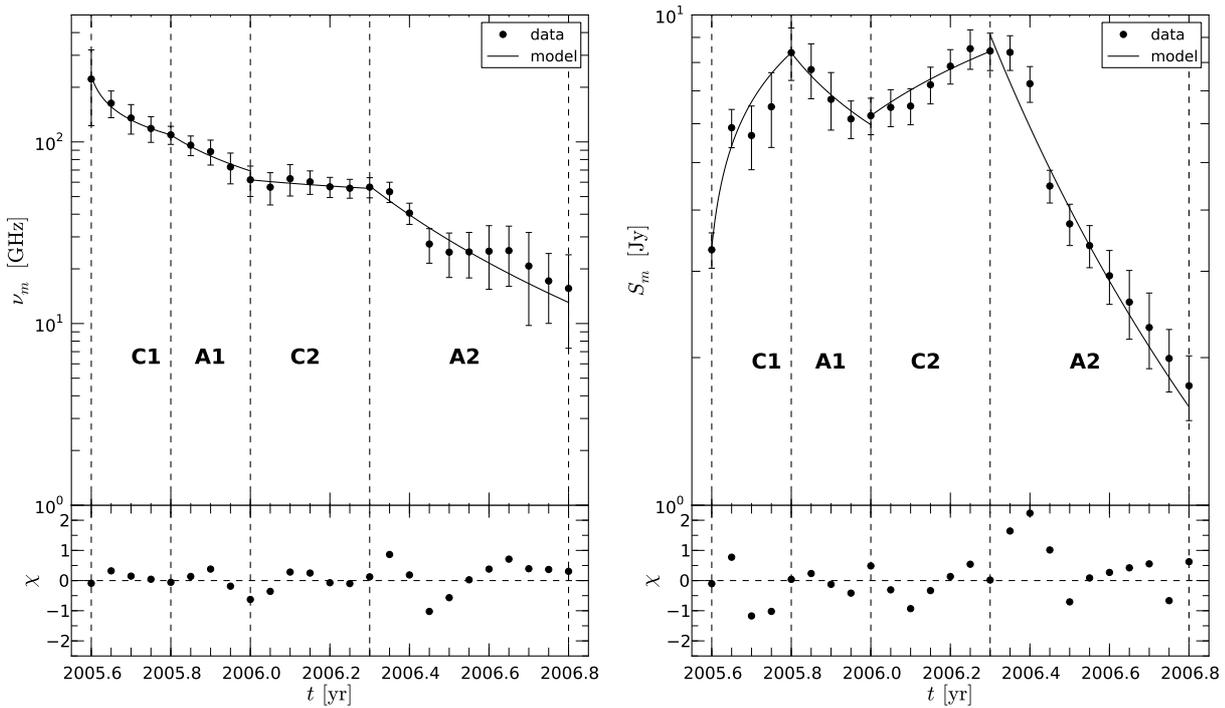} 
\caption{Temporal spectral evolution of the 2006 radio flare in CTA\,102 left: turnover frequency; right: turnover flux density. The lower
panels show the residuum for the fits $\chi=\left(x_{\mathrm{obs}}-x_{\mathrm{model}}\right)/\Delta x$. 
The dashed black lines correspond to the time labels in Figure \ref{numsm} and indicate the extrema in the evolution.} 
\label{finalfig} 
\end{figure*}

\begin{table}
\caption{Best fit values for spectral evolution modeling of the 2006 radio flare in CTA\,102, parameters $b$, $d$, $s$, $k$ $r$ and $t$.}
\label{finaltab}
\begin{tabular}{c c c c }
\hline \hline
 & 2005.60$-$2005.95 & 2005.95$-$2006.30 & 2006.30$-$2006.80\\
 & C1A1 & C2 & A2\\
\hline
\noalign{\smallskip}
$b$ &\rule[-0.25cm]{0cm}{0.4cm} 1.0$^{+0.08}_{a}$ 	& 1.35$^{+0.65}_{-0.35}$	&1.7$\pm$0.2\\

$d$ &\rule[-0.25cm]{0cm}{0.4cm} 0.2$\pm$0.02 	&$-$0.1$\pm$0.03		&$-$0.2$^{+0.08}_{-0.05}$\\

$s$ &\rule[-0.25cm]{0cm}{0.4cm} 2.1 	& 2.0		& 2.4\\

$k$ &\rule[-0.25cm]{0cm}{0.4cm} 2.7$\pm$0.14 		& not fitted 		& 4.7$\pm$0.4\\

$r$& \rule[-0.25cm]{0cm}{0.4cm} 0.60$\pm$0.03 		& 0.35$\pm$0.02	& 0.90$\pm$0.07\\

$t$& \rule[-0.25cm]{0cm}{0.4cm}  0.02$\pm$0.01 	& \multicolumn{2}{c}{not fitted}\\
\hline
\end{tabular}
\begin{list}{}{}
\item[$^{a}$] value hits lower boundary.
\end{list}
\end{table}

\subsubsection{Spectral evolution before 2006.0}
\label{synch}
The spectral evolution until 2006.0 followed approximately the standard evolution described by the shock-in-jet 
model. However, we have not found evidence for a plateau phase of turnover flux density, $S_{m}$, {neither after the first Compton stage nor after the  second Compton-like stage.}  Therefore, we excluded the synchrotron stage from our modeling using a direct 
transition from the Compton to the adiabatic one. The result of this modeling is presented in Table~\ref{finaltab} 
as model C1A1.

\subsubsection{Spectral evolution between 2006.0 and 2006.3}
\label{altermod}
The second peak in the $S_m - \nu_m$ plane shows a similar behavior to a Compton stage, as stated above. Therefore, we applied the 
equations of the Compton stage to the spectral evolution between 2006.0 and 2006.3.
Since the Compton stage can be explained within a 4-dimensional parameter space 
(note that it does not depend on parameter $k$, Eqs.~\ref{an1} $-$ \ref{aq3}), we selected carefully a physically meaningful 
combination of parameters.

Boundaries for the slope of the relativistic electron distribution, $s$, can be derived by using the evolution of the 
optically thin spectral index of the emission  $\left(s=1-2\alpha_0\right)$ between 2006.0 and 2006.3. This evolution is shown in 
Fig.~\ref{alpha}. We obtained values of $s_\mathrm{min}=1.4$ and $s_\mathrm{max}=2$ from the optically thin 
spectral indeces $\alpha_0=-0.20$ and $\alpha_0=-0.47$, respectively.

\subsubsection{Spectral evolution after 2006.3}
\label{evo2006.3}
The spectral evolution after 2006.3 shows the typical behavior of an adiabatic loss stage. 
Again, we used the evolution of the optically thin spectral index, $\alpha_0$, shown in 
Fig.~\ref{alpha},to derive limits for the parameter $s$. The obtained values are 
$s_\mathrm{min}=1.6$ and $s_\mathrm{max}=2.5$. This stage could be divided into two 
substages (before and after 2006.45) but the limited time sampling (with only three data points 
in the first part) forced us to perform the spectral fitting to the whole adiabatic stage.

\subsection{Final model and error analysis}

Table~\ref{finaltab} lists the best fits and errors of the different stages. 
In the first column, we show the values for C1A1. Since we have not found evidence for a 
synchrotron stage, we assumed a direct transition between a Compton and an adiabatic stage.

A study of the uncertainties of the spectral parameters as in \cite{Lampton:1976p2319} is not
suitable due to the small number of data points and the strong mathematical interdependence
of the parameters. This yields to mathematically correct, but non-physical solutions. For the same
reason, we did not perform an analysis of the large 5 or 6-dimensional parameter space.
To provide first-order error estimates we followed this approach:
The set of parameters derived minimizes the $\chi^{2}$-distribution in the 
parameter space, so we investigated the stability of this point in the parameter space. 
This error analysis is based on the variation of the $\chi^2$ for a given parameter sweep 
within the listed boundaries (Table~\ref{boundaries}), while keeping the others fixed. 
The final values for the uncertainties were obtained by calculating the 68\% probability 
values, assuming a normal distribution for the values of the parameters. Note that the $\chi^2$ 
distribution is not always symmetric around the minimum value, and this leads to lower and upper 
limits for the error estimates. Fig.~\ref{finalfig} shows the result of our spectral modeling.

\subsection{Modeled spectra and light curves}
\label{modelspec}
\begin{figure}[h!]
\resizebox{\hsize}{!}{\includegraphics{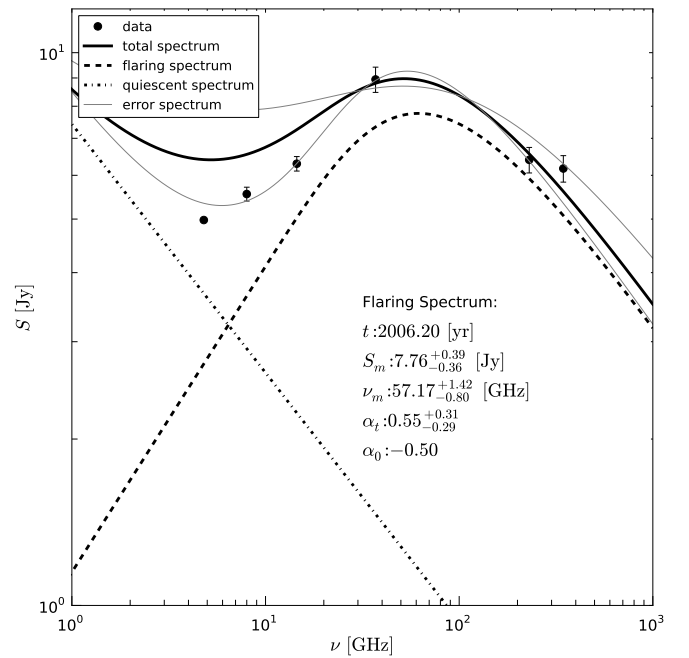}} 
\caption{Modeled spectrum for the 2006.20 observations.The dashed-dotted line corresponds to the quiescent spectrum, the 
dashed line to the flaring spectrum, and the solid black line to the total spectrum. The gray lines indicate the uncertainties in the calculation of the spectrum ({for} more details see Sect. \ref{modelspec}.} 
\label{2006.20m} 
\end{figure}

With the derived parameters and their error estimates we calculated the evolution of flaring spectrum and compared it to the 
observed flux density values at any given epoch. For each time step during the flaring activity the turnover frequency and 
the turnover flux density can be thus calculated. The optically thin spectral index follows from the assigned spectral slope, $s$. 
The calculation of the optically thick spectral index, $\alpha_t$, can be performed following the approach of \citet{Turler:2000p1}, setting $\alpha_t\sim f_3/n_3$ (see Eqs. \ref{f3} and \ref{n3}).
This leads to a set of four spectral parameters $(S_m\,,\nu_m\,,\alpha_0\,,\alpha_t)$ from which the shape of the flaring spectrum can 
be derived by using Eq.\ref{snu}. By incorporating the quiescent spectrum, the total spectrum can be 
computed. From the uncertainties of the parameters ($b$, $s$, $k$, $d$, and $r$) and the quiescent spectrum, the lower and upper boundaries for the modeled spectra can be obtained. Fig. \ref{2006.20m} shows the modeled spectrum for the 2006.2 observations as an example.
The calculation of the optically thick spectral index, using $\alpha_t\sim f_3/n_3$ led to values 
which reproduce well high frequency part of the spectrum but not the low frequency one 
(see Fig.\ref{2006.20m}). {The discrepancy in the optically thick part of the spectrum was probably caused by
the quiescent contribution, which is known to vary slightly over time (see variations in the flux density for $t<1995$ Figure \ref{archival}).}

\begin{figure}[h!]
\resizebox{\hsize}{!}{\includegraphics{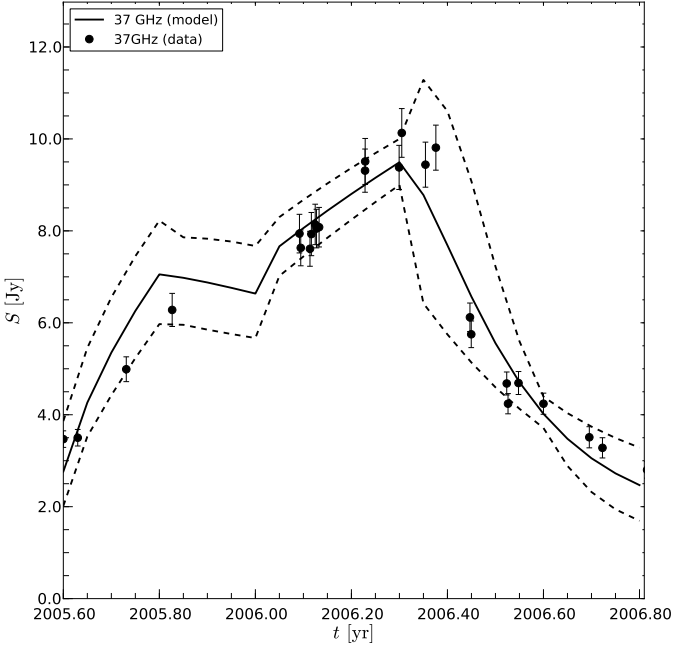}} 
\caption{Modeled and observed 37\,$\mathrm{GHz}$ light curve. The black solid line corresponds to the calculated model and the dashed lines indicate the uncertainties in the calculation of the modeled light curve ({for} more details see text).} 
\label{mlc} 
\end{figure}

Once the spectra are calculated, it is straightforward to obtain the modeled light curves at a given frequency. 
The comparison between the modeled and the observed 37\,$\mathrm{GHz}$ light curve is presented in Fig.~\ref{mlc}. Although similar 
plots can be obtained for other frequencies, it is meaningful to select a frequency at which the observed emission is mainly generated 
by the interaction of the traveling shock with the underlying flow. Moreover, the dense sampling of the 37\,$\mathrm{GHz}$ 
observations provided a more complete picture of the evolution of the flaring event. Taking this consideration into account, we picked 
the 37\,$\mathrm{GHz}$ light curve as the best possible comparison. The observed data points fall well within the range of modeled 
light curve. At the beginning of the flare, the 37\,$\mathrm{GHz}$ flux density was still in its quiet state and therefore the main 
uncertainty for this time is due to the uncertainties in the quiescent spectrum. The error band presented here corresponds only to the 
flaring state, which was the dominant contribution to the 37\,$\mathrm{GHz}$ flux density after 2005.60, leading to broader error 
bands from this epoch and on. 

\subsection{Rejected solutions} 
As already mentioned, the solutions for the spectral fitting were embedded in a 4- or 5-dimensional parameter space. Some of 
those can be highly degenerate given the small number of data points. When fitting the spectral evolution, a deep study 
of the parameter space was performed, including all possible combinations of the parameters, (i.e., variation of single, pairs and 
triplets of parameters) and simultaneously fitted stages, (one-, two- and three-stage fits). Within this study we found families 
of solutions with $\chi^2$ values similar to the presented final set of parameters, but with unphysical values. After a sanity check, those fits were discarded. The meaningful set of solutions is presented in Table~\ref{finaltab} and 
discussed in Sect.~\ref{disc}.

\begin{figure*}  
\includegraphics[width=17cm]{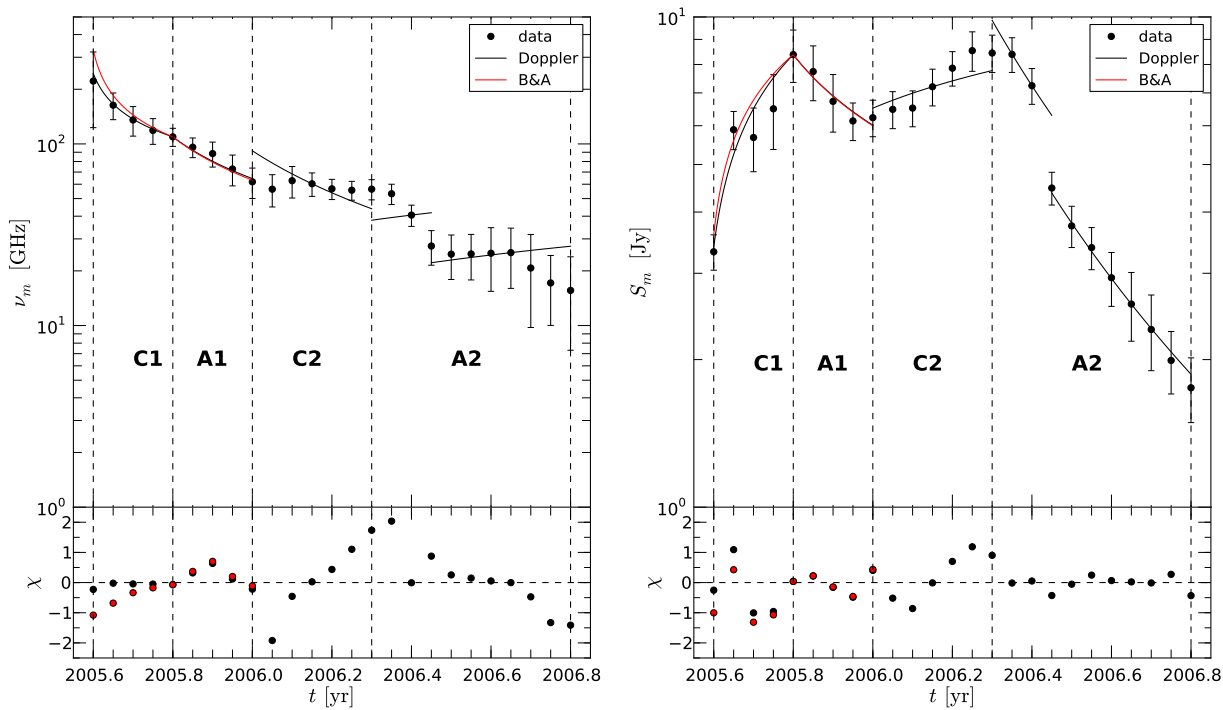} 
\caption{The temporal evolution of the 2006 radio flare in CTA\,102 modeled with varying Doppler factor (black solid line) {for $t>2006.0$} and with the modifications of \citet{Bjornsson:2000p32} for the first two stages (red solid line). Both models fail to describe the 2006 flare (compare to Fig. \ref{finalfig}).; left: turnover frequency; right: turnover flux density. The lower panels show the residuum for the fits $\chi=\left(x_{\mathrm{obs}}-x_{\mathrm{model}}\right)/\Delta x$. 
The dashed black lines correspond to the time labels in Figure \ref{numsm} and indicate the extrema in the evolution. } 
\label{reject} 
\end{figure*}

\subsubsection{The geometrical model}
\label{gm}
\citet{Stevens:1996p449} found a similar double hump in the $\nu_{m}-S_{m}$ plane for 3C\,345.
They assumed that this behavior could be due to changes of the viewing angle, expressed in
a variation of the Doppler factor, $D$, along the jet. Therefore we used as an alternative
to the previous modeling of the second hump in the $\nu_m-S_m$ plane, 
a purely geometrical approach. {Such an approach could easily explain the variation in the observed turnover flux density $\left(S^{\prime}_{m}\propto D^{3-\alpha_{0}}S_{m}\right)$ while the observed turnover frequency kept a nearly constant value $\left(\nu^{\prime}_{m}\propto D\nu_{m}\right)$.}
It was assumed that the deviation from the standard shock-in-jet model around 
2005.95 {was} caused by changes in the 
evolution of the Doppler factor $D=\Gamma^{-1}\left(1-\beta\cos\vartheta\right)^{-1}$ during 
the final adiabatic loss stage. 
The remaining parameters were taken from
the best fit to the evolution until 2006.0 (first column in Table~\ref{finaltab}, model C1A1) and 
only the parameter $d$ was allowed to vary.

The result of our calculations for the different models is presented in Fig.~\ref{reject} and 
Table~\ref{gmodres}.
The figure shows the fits separating the different stages observed after 2006.0 
(Figs.~\ref{numsm} and \ref{numsmtime}). We recall that these stages are the increase of the 
flux density between 2006.0 and 2006.3 and the decrease in flux density and frequency from 2006.3
until 2006.8.

\begin{table}[h!]
\caption{Results of the geometrical model for parameter $d$.}
\label{gmodres}
\centering
\small{
\begin{tabular}{c c c }
\hline\hline
time & type & $d$\\
\hline
2006.0$-$2006.8 &3-stage &$-$0.04\\
2006.0$-$2006.5 &2-stage &$-$0.30\\
2006.5$-$2006.8 &2-stage &$-$1.24\\
2006.0$-$2006.3 &1-stage &$-$0.41\\
2006.3$-$2006.5 &1-stage &$-$1.33\\
2006.5$-$2006.8 &1-stage &$-$1.27\\	 
\hline
\end{tabular}}
\end{table}

It is obvious that a simple change in the evolution of the Doppler factor, expressed by the exponent 
$d$, $\left(D\propto R^{-d(t)}\right)$, can not explain the observed temporal evolution of the turnover frequency and 
turnover flux density.  

\subsubsection{Applying the modified compton stage}\label{nolg}
\begin{figure}[h!]
\resizebox{\hsize}{!}{\includegraphics{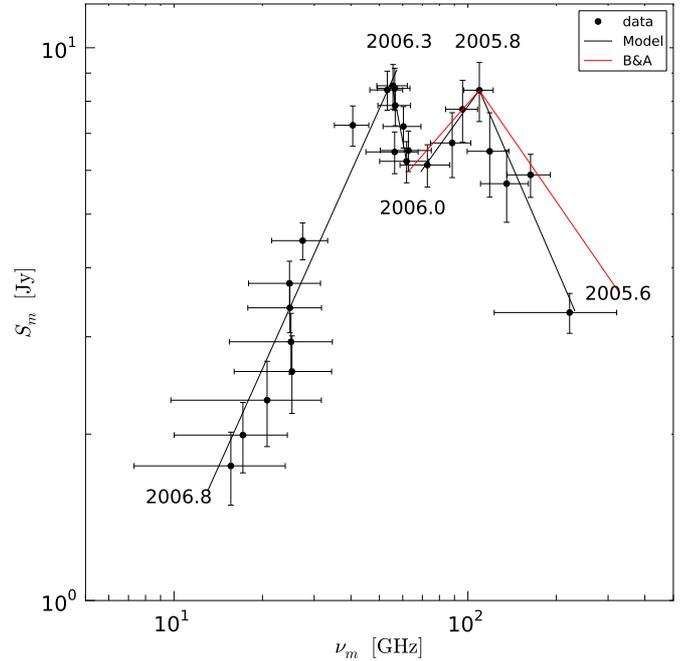}} 
\caption{The 2006 radio flare in the turnover frequency - turnover flux density plane. The time labels indicate the time evolution 
and the temporal position of local and global extrema in the spectral evolution. The black line corresponds to the final model presented in Table\ref{finaltab} \citep[based on][]{Marscher:1985p50} and the red one to Table \ref{tabbj} \citep[based on][]{Bjornsson:2000p32}. } 
\label{numsmfit} 
\end{figure}

{As mentioned in Sect. \ref{theory}, \citet{Bjornsson:2000p32} reviewed \citet{Marscher:1985p50} assumption
for the Compton stage and modified the equations for the evolution of the turnover frequency, $\nu_{m}$, and the 
turnover flux density, $S_{m}$. We applied this model to the first two stages of the 2006 flare in CTA\,102 (Model C1A1), using 
the expression as in Sect.~\ref{shockinjet} for the time evolution (Eq.~\ref{t1}) and $d=0$. The 
result of the modeling is shown in Fig. \ref{numsmfit}, Fig.\ref{reject} and Table \ref{tabbj}.}

\begin{table}[h!]
\caption{Spectral parameters for C1A1 using \citet{Bjornsson:2000p32}}
\label{tabbj}
\centering
\begin{tabular}{c c c c c c}
\hline \hline
$b$ & $s$ & $k$ & $d$ & $r$ &$t$\\
\hline
1.0 & 1.6 & 2.9 & 0& 0.6 & 0.02 \\
\hline
\end{tabular}
\end{table}

The result of the fitting shows that the approach of \citet{Bjornsson:2000p32} cannot explain 
the steep rise of the turnover flux density with decreasing turnover frequency in the first 
stage of the flare (see solid red line in Fig. \ref{numsmfit}). Following the authors, this could be an indication that inverse Compton 
scattering is not the dominant energy loss mechanism during the rising stage 
\citep{Bjornsson:2000p32}. This aspect will be discussed in Sect.~\ref{disc}.

\section{Discussion}
\label{disc}
{Our analysis of the strong radio flare observed in CTA~102 around 2006, shows that its behavior 
until epoch 2006.0, i.e., the first hump in the $\nu_m-S_m$ plane, can be well modeled by a Compton and adiabatic 
stage within the standard shock-in-jet model 
\citep{Marscher:1985p50} (see Sect. \ref{synch}). We found no evidence of a synchrotron stage between the Compton 
and the adiabatic stages during the first part of the evolution. This behavior was similar to the one found in another prominent blazar 3C\,345 \citep{Lobanov:1999p2299}, where a bright jet component also appeared to proceed from the Compton stage to the adiabatic stage, with the synchrotron stage ruled out by combination of the observed spectral and kinematic evolution of this component. The flare in CTA\,102 shows a second hump in the $\nu_m-S_m$ 
plane (see Fig.\ref{numsmfit}) after 2006.0, that cannot be explained within this model.
The evolution between 2006.0 and 2006.3 seems to follow the predicted evolution of a Compton stage, i.e., 
increasing turnover flux density and decreasing the turnover frequency. Based on this apparent behavior we applied 
the equations of the Compton stage to this second hump (see Sect. \ref{altermod}). After epoch 2006.3, the evolution 
shows a decrease in both turnover values, which can be interpreted as an adiabatic stage. Thus, we used the 
appropriate equations for this stage to fit the data points (see Sect. \ref{evo2006.3}). This 
approach led to a set of exponents which describe the evolution of the magnetic field, $B$, the Doppler factor, 
$D$, the spectral slope of the electron distribution, $s$, the 
normalization coefficient of the relativistic electron distribution, $K$, and the jet radius, $R$. The 
results of our modeling are summarized in Fig.~\ref{finalfig}. Our hypothesis is that this behavior 
can be interpreted in terms of the interaction between a traveling shock and a standing shock wave in an 
over-pressured jet. A pure geometrical model failed to explain the observed behavior 
(see Sect. \ref{gm}). We also tried to fit the evolution of the flare using the modification of the shock-in-jet
model by \citet{Bjornsson:2000p32}, but this also failed. However, the failure of this model, which takes into 
account multiple Compton scattering, could have further implications in our understanding of flaring events in AGN 
jets, as discussed at the end of this section.}

The formation of a standing shock can be described in the following way: The unbalance between 
the jet pressure and the pressure of the ambient medium at the jet nozzle leads to an opening of the jet. 
Due to the conservation laws of hydrodynamics, this opening results in a decrease of the density, the 
pressure and the magnetic field intensity in the jet. The finite speed of the sound waves in the jet is 
responsible for an over-expansion followed by a re-collimation of the jet that gives rise to the 
formation of a shock. During this collimation process the jet radius decreases and the shock 
leads to an increase of the pressure, density, and magnetic field intensity. Again, the finite speed of 
the sound waves is responsible for a over-collimation of the jet. This interplay between over-expansion 
and over-collimation leads to the picture of a pinching flow, i.e., a continuous change of the width along 
the jet axis, in contrast to conical jets. The intrinsic physical parameters (pressure, density, and 
magnetic field) along a pinching jet show a sequence of local maxima and minima 
\citep{Daly:1988p3,Falle:1991p496}.

\begin{figure}[h!]
\resizebox{\hsize}{!}{\includegraphics{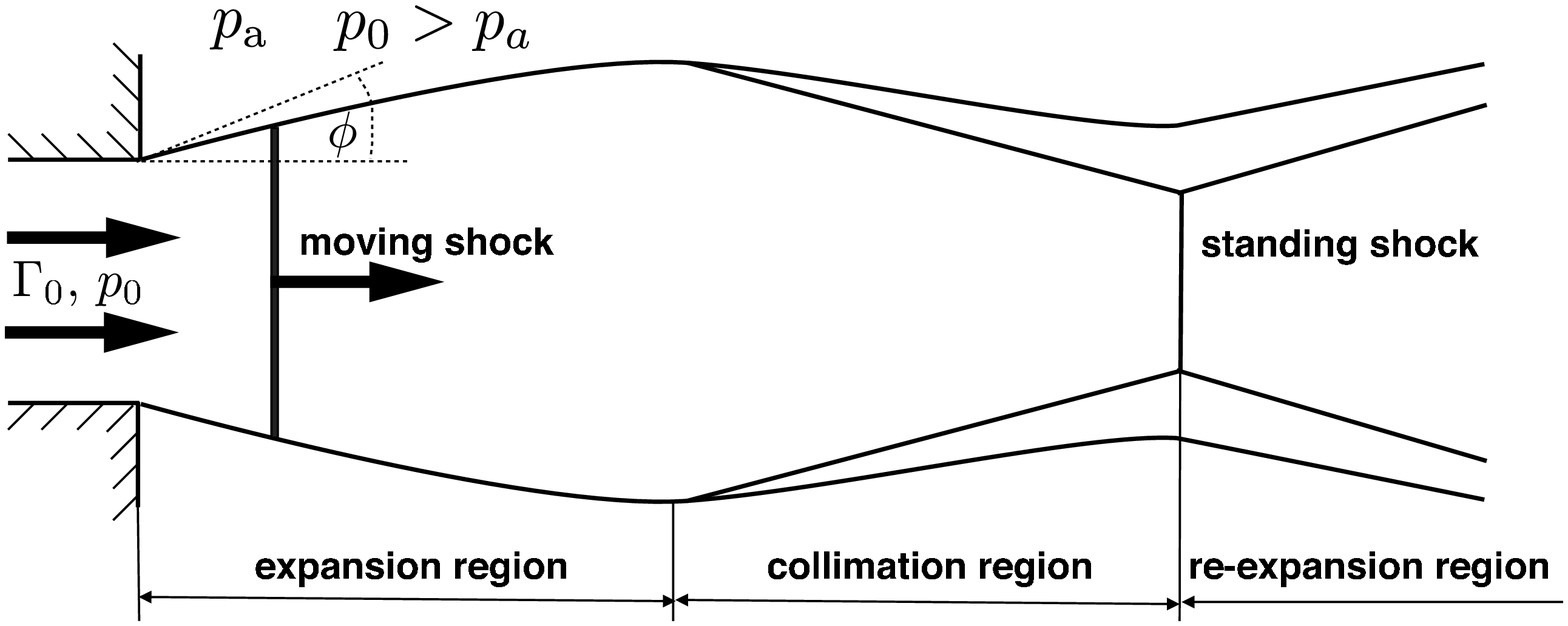}} 
\caption{Sketch of an over-pressured jet with indicated characteristic regions \citep[adopted from][]{Daly:1988p3}. } 
\label{jetsketch} 
\end{figure}

Summarizing, an over-pressured jet can be 
divided into three regions, i) the expansion region, i.e., continuous increase of the jet radius, ii) the 
collimation region, i. e., decrease of the jet radius and formation of the collimation shock, and iii) 
the re-expansion region (see Fig.~\ref{jetsketch}). In such a scenario, enhancements of emission can be 
produced by the interaction between traveling 
and standing (re-collimation) shocks \citep{Gomez:1997p649}. In the following, the evolution
of the travelling shock in these regions is described and our results are put in context.

\subsection{The expansion region}
\label{expan}
A relativistic shock propagating through this region accelerates particles at the shock 
front. These particles travel behind the shock and suffer different energy loss mechanisms, 
depending on their energy. The resulting evolution of the turnover frequency and turnover flux density is 
explained by the shock-in-jet model \citep{Marscher:1985p50} under certain assumptions.

The parameters derived for the time between 2005.6 and {2006.0 can be associated with this region, including a Compton and an adiabatic stage}. The 
expansion of the jet is parametrized by $r=0.60$, which differs from the value expected for a conical jet $r=1$. 
The non-conical behavior could be due to acceleration of the flow \citep{Marscher:1980p1589}. 
This acceleration should be expressed by $-r \cdot d>0$ ($D\propto L^{-rd}$). However, from our results, this value is 
negative $-r \cdot d = -0.12$, but very small, i.e.,
compatible with no changes in the Doppler factor. Thus, we cannot confirm this point.  

For the evolution of the magnetic field with distance we derive a value of $b=1.0$, 
indicating that the magnetic field could be basically toroidal in this region. The injected spectral slope 
for the relativistic electron distribution $s=2.1$ leads to an  
optically thin spectral index $\alpha_0=-0.55$. A decrease in the density can be deduced from 
$-r \cdot k=-1.6$, which corresponds to the evolution of the normalization coefficient of the 
relativistic electron distribution, $K$.

The parameter $t_\mathrm{off}$ corresponds to the time difference between the onset of the Compton stage 
and the first detection of the flare. From the value of $t_\mathrm{off}=0.02\,\mathrm{yr}$ 
together with independently obtained values for the 
viewing angle, $\vartheta=2.6^\circ$, and the apparent speed of the VLBI component ejected by the 2006 
flare, $\beta_\mathrm{app}=17\,c$ \citep{Jorstad:2005p64,Fromm:2010p1041}, we calculated the displacement 
between the onset and the detection of the flare to be $\Delta r= 3.5\,\mathrm{pc}$. 

\subsection{The collimation region}
After the recollimation region, at the position of the hypothetical standing shock, the local increase 
in density, pressure and magnetic field should generate an increase in the emission. The interaction 
between a travelling 
and a standing shock would further enhance the emission \citep{Gomez:1997p649}. 
Furthermore, the standing shock would be dragged downstream by the traveling shock and 
re-established after a certain time at its initial position \citep{Gomez:1997p649,Mimica:2009p42}. 
We compare here our results with this scenario.

Since the evolution of the turnover frequency and turnover flux density between 2006.0 and 2006.3 
showed Compton-stage-like behavior, i.e., decreasing turnover frequency and increasing flux density, 
we used the equations of the Compton stage to derive the possible evolution of the physical 
parameters (model C2).  In this region a slower rate of jet expansion is found ($r=0.35$). During this stage,
the Doppler factor seems to be constant with distance, $-r \cdot d=0.035$. In the context of the hypothetical
shock-shock interaction, acceleration of the 
flow close to the axis is expected down to the discontinuity of the stading shock, 
where sudden deceleration would occur \citep[see, e.g.,][]{Perucho:2007p9}. Thus, it is difficult to assess whether the 
Doppler factor should increase or decrease in the whole region. 

The magnetic field intensity decreases with an exponent 
$b=1.35$, implying that the geometry of the magnetic field has changed, with contributions of non-toroidal components, but showing no hints of 
magnetic field enhancement. The parameter $s$, giving the spectral slope 
of the relativistic electron distribution changes to $s=2$, which gives an 
optically thin spectral index $\alpha_0=-0.5$. 

The set of parameters derived for this time interval do not reflect the expected physical conditions of a 
traveling$-$standing shock interaction, other than a slight flattening of 
the spectral slope (from possible refreshment of particles). Nevertheless, the shock-shock scenario could hardly be reproduced by a 
one-dimensional model. Numerical simulations should be performed in order to study this hypothesis in 
detail. Another possibility is that the reason for the second peak is attached to the injection of a second 
shock from the basis of the jet. This is not observed, though. 

\subsection{The re-expansion region}
After the re-collimation process, the jet re-expands, i.e., the jet radius increases again. 
In principle, the position of the re-collimation shock can be regarded as a ``new'' nozzle from which the 
fluid emerges. Therefore, when the shock front reaches this region, the expected evolution is, again, 
that predicted by the shock-in-jet model. 

The evolution between 2006.3 and 2006.8 is identified within our hypothesis with the re-expansion region. 
Thus, the equations for an adiabatic loss stage were applied to the evolution turnover frequency and 
turnover flux density. 

The opening of the jet is clearly apparent at this stage $r=0.90$. This opening should produce a decrease 
in density, which translated into smaller values of the 
parameter $-r\cdot k$, $\left(K\propto L^{-r\,k}\right)$. From the fits, we derive $-r \cdot k=-4.2$, 
confirming a decay in the density. The magnetic field falls with $b=1.7$, which shows 
again that the geometry of the field changes from a purely toroidal 
to a mixed structure with the distance. The values for $-r \cdot d = 0.18$ reveal an 
acceleration of the flow, which can naturally arise during the expansion of the jet. The spectral slope of $s=2.4$ 
translates into an optically thin spectral index of $\alpha_0=-0.7$.

\subsection{Spectral slopes and optically thin spectral indices}

The variation of the fitted optically thin spectral index, $\alpha_{0,\mathrm{f}}$, 
(see upper panel in Fig.~\ref{alpha}) is an indication
of the aging of the relativistic electron distribution due to the different energy loss mechanisms 
during the evolution of the flare. 
The model presented by \citet{Marscher:1985p50} did not include such an aging of the relativistic electron 
distribution. Despite this restriction, we could model the qualitative 
behavior of the evolution via the different values obtained for parameter $s$ in the different fitted 
stages: For the expansion region we obtained $\alpha_{0,\mathrm{m}}=-0.55$, a slightly flatter 
spectral index for the collimation region  $\alpha_{0,\mathrm{m}}=-0.50$ and a steeper value for the re-expansion 
$\alpha_{0,\mathrm{m}}=-0.7$.

\subsection{The influence of the interpolation and the quiescent spectrum}
We applied several interpolation steps together with different quiescent spectral parameters to the 
analysis of the light curves in order to test their influence on the study of the evolution of the peak 
parameters in the turnover frequency$-$turnover flux density ($\nu_m-S_m$) plane. All the tests showed a 
second hump in the $\nu_m-S_m$ plane. From the tests, the positional shifts (in time of appearance, 
turnover frequency and turnover flux density) of the second hump were calculated: The hump appeared at 
$t=2006.30\pm0.05\,\mathrm{yr}$ at a turnover frequency $\nu_m=56^{+5}_{-3}\,\mathrm{GHz}$ and turnover 
flux density $S_m=8.5^{+0.1}_{-0.9}\,\mathrm{Jy}$. 

These tests proved that the increase of the turnover frequency and turnover flux density around 
2006.3 is not an artifact generated by the interpolation and/or the choice of the quiescent spectrum.

\subsection{Adiabatic versus non-adiabatic shock} \label{advsnad}

 \citet{Bjornsson:2000p32} suggested that the steep slopes in the rising region of the $\nu_m$-$S_m$- 
plane, which is generally identified as the Compton stage in the shock-in-jet model, could not be due to 
Compton radiation, but to other processes such as isotropization of the electron distribution or even to 
the non-adiabatic nature of the shock. 

The slope we obtained for this time-interval in the $\nu_m$-$S_m$ plane is too steep to 
be reproduced by the model presented in \citet{Bjornsson:2000p32} (see Sect.~\ref{nolg}). Thus, the 
non-adiabatic nature of the first stages of evolution of shocks in extragalactic jets remains an open 
issue on the basis of our limited data set.

\section{Summary and conclusions}
{In this work we present the analysis of the 2006 flare in CTA\,102 in the $\mathrm{cm}-\mathrm{mm}$ regime. The obtained evolution could be well described, up to a certain point ($t<2006.0$), with the standard shock-in-jet model \citep{Marscher:1985p50}. In order to model the further evolution of the flare we proposed a second Compton-like stage and a second adiabatic stage. We derived the evolution of the physical parameters of the jet and the flare and we performed a parameter space analysis to obtain the uncertainties of the values obtained. From the modeled 
parameters, together with their uncertainties, the theoretical light curves and spectra were computed. The result was shown to 
be in fair agreement with the observations.}

However, the shock-in-jet model is not able to reproduce the second peak of the double-hump structure found
in the evolution of the peak flux - peak frequency plane in a consistent way within our hypothesis of a shock-shock
interaction.This is surely due to the complexity of the situation and does not rule out our hypothesis. Numerical simulations 
will be performed to test it. The goodness of the fit in this region may be due basically to the lack
of observational data as compared to the number of parameters implied in the modeling. 

Moreover, our results show that some of the basic assumptions
of the present models should be reviewed. In particular,  we found that the first stage of the studied 
evolution could be incompatible with an adiabatic behavior if the model of \citet{Bjornsson:2000p32} is correct. 

We plan to improve our results by performing a similar analysis to multifrequency VLBI observations. 
Future work on the spectral evolution of components in jets, 
especially the interaction between traveling shocks and re-collimation shocks, should include numerical 
simulations to help 
to understand this non-linear phenomenon, and improve the strong limitations of the assumptions required by 
the shock-in-jet and other analytical models. The analysis of multifrequency VLBI observations during the 2006 radio
flare in CTA\,102 will be presented in paper II.

\begin{acknowledgements}
C.M.F. was supported for this research through a stipend from the International Max Planck Research School (IMPRS) for Astronomy and 
Astrophysics at the Universities of Bonn and Cologne. Part of this work was supported by the COST Action MP0905 Black Hole in a 
violent Universe. C.M.F. was supported during his Diploma Thesis by K\"olner Gymnasial$-$ und Studienstiftung.
C.M.F. thanks the Departament d'Astronomia i Astrof\'\i sica of the University Val\`encia for its hospitality.\\
M.P. acknowledges financial support from the Spanish ``Ministerio de Ciencia e Innovaci\'on'' (MICINN) grants 
AYA2010-21322-C03-01, AYA2010-21097-C03-01 and CONSOLIDER2007-00050, and from the ``Generalitat Valenciana'' grant 
``PROMETEO-2009-103''. M.P. acknowledges support from MICINN through a ``Juan de la Cierva'' contract.\\
E.R.  acknowledges partial support by the Spanish MICINN through grant AYA2009-13036-C02-02.\\
Part of this work was carried out while T.S. was a research fellow of the Alexander von Humboldt Foundation.\\ 
This work is based on observations with the radio telescope of the university of Michigan, MI, USA, the  Mets\"ahovi radio 
telescope of the university of Helsinki, Finland and Submillimeter Array (SMA) of the Smithsonian Astrophysical Observatory, 
Cambridge, MA, USA. The operation of UMRAO is made possible by funds from the NSF and from the university of Michigan. The 
Submillimeter Array is a joint project between the Smithsonian Astrophysical Observatory and the Academia Sinica Institute of 
Astronomy and Astrophysics and is funded by the Smithsonian Institution and the Academia Sinica. The Mets\"ahovi team acknowledges 
the support from the Academy of Finland to our observing projects
We thank Maria Massi and Petar Mimica for fruitful discussions and comments.
\end{acknowledgements}

\bibliographystyle{aa} 
\bibliography{16857}

\end{document}